**Thermal excitation of flexoelectricity in silicon**

Lingtong Lv,[1] Qianqian Ma,[1] Kailu Wang,[1] Xin Wen,[1,2*] and Shengping Shen[1,†]

[1] State Key Laboratory for Strength and Vibration of Mechanical Structures, Xi'an Jiaotong University, Xi'an 710049, China
[2] Institut Catala de Nanociencia i Nanotecnologia (ICN2), CSIC and The Barcelona Institute of Nanoscience and Technology (BIST), Campus Universitat Autónoma de Barcelona, Bellaterra, Catalonia

**ABSTRACT**. Flexoelectricity, an electromechanical coupling between strain gradient and polarization, offers a promising dimension to enrich silicon-based devices. Although the flexoelectricity of silicon is known, some fundamental aspects remain ambiguous, such as the discrepancy between experimental results and theoretical predictions, the influence of doping concentration, and the role of the bandgap. Here, we measured the flexoelectricity of intrinsic and heavily doped Si over the temperature range of -50 °C~200 °C. The flexoelectric coefficient is of ~2.6 µC/m and barely varies with temperature in doped silicon, while in intrinsic silicon it varies by nearly two orders of magnitude from ~15.2 nC/m to 1.8 µC/m as temperature increases. We show that their different temperature dependencies correspond to the temperature-insensitive donor ionization in doped silicon and the temperature-sensitive intrinsic excitation in intrinsic silicon, with the latter captured by a quantitative relationship between flexoelectricity, temperature and bandgap. Furthermore, similar experimental results on germanium (Ge) suggest the universality of this relationship in first-generation semiconductors. These findings would offer valuable reference for developing Si-based electromechanical devices, as well as understanding the strain-gradient effects on semiconductor band structures (flexoelectronics).

Silicon-based devices, such as diodes [1,2], transistors [3,4], and solar cells [5], lie at the heart of modern semiconductor technology. Electromechanical coupling, a functionality enabling the conversion between mechanical and electrical energy, may offer a dimension for enriching silicon-based device applications. Unfortunately, silicon does not generate electricity under pressure due to its centrosymmetric crystal structure (lack of piezoelectricity), limiting its applicability in this context [6]. This void can be supplemented by flexoelectricity, a coupling that links polarization to strain gradients in all insulators irrespective of their structural symmetry [7–9], which has been recently demonstrated in semiconductors (including Si) as well [10–25].

Research on silicon flexoelectricity dates back to the ab-initio calculations by Hong & Vanderbilt [26] who reported a longitudinal flexoelectric coefficient $\mu_{11}$ of 1.3 nC/m and a flexocoupling coefficient $f_{11}$ ($\mu_{11}$ normalized by dielectric constant) of 12 V. Stengel [27,28] later reported comparable results and completed the transverse and shear components ($\mu_{13}$, $\mu_{44}$). Mizzi & Marks [29] further took account the surface contribution and reported an effective $f_{13}^{eff}$ of ~6 V (corresponding to a $\mu_{13}^{eff}$ of ~0.7 nC/m) under a beam-bending configuration. These theoretical predictions, however, deviate significantly from experimental results. Preliminary measurements by Narvaez [30], the only direct flexoelectric characterization of Si to our knowledge, reported a much larger $\mu_{13}^{eff}$ in the range of 40~80 nC/m for phosphorus-doped Si. Narvaez tentatively attributed this enhancement to the barrier layer mechanism occurring at the metal-semiconductor interfaces [22], an effect that had not been considered in previous calculations [26–29]. Yet the observed weak doping dependence is not what would typically be expected from this mechanism, via which doping often alters flexoelectricity by orders of magnitude [22,23,31,32]. The discrepancy between the experiment and calculations—and to what extent the doping level matters—remain open questions that warrant systematic revisitation, especially given the surge of interest in the flexoelectric modulation for silicon electronics and photovoltaics [11–21].

In this study, we measured the effective flexoelectric coefficient $\mu_{13}^{eff}$ in both intrinsic and heavily doped (Arsenic (As) doped, ~$10^{18}$ cm$^{-3}$) silicon, and found a doping-induced enhancement by two orders of magnitude (-50 °C~200 °C). Notably, a comparable enhancement was also observed in undoped silicon as the temperature increased. We demonstrate, by theoretically linking flexoelectricity and bandgap, that this thermally enhanced flexoelectricity is a result of the intrinsic excitation of free carriers.

We performed the flexoelectric characterizations, firstly at room temperature, on gold-electroded silicon beams using a standard three-point bending method, where a static force was applied to fix the sample at place while an oscillating force was applied to induce flexoelectric charge (see Section 2 in Supplemental Material [33]). Although the magnitude of the static force has been found to affect the flexoelectric coefficient via polar ferroelastic domain walls [34,35], silicon is not ferroelastic and such dependence was not observed in our measurement (see Fig. S2 in Supplemental Material [33]). Fig. 1a and 1b show the synchronous sinusoidal charge output induced by the sinusoidal bending deformation for heavily doped and intrinsic silicon respectively. The slope of the linear relationship between flexoelectric polarization and strain gradients, as shown in Fig. 1c, represents the effective flexoelectric coefficient ($\mu_{13}^{eff}$), which were measured to be 2.58 µC/m for heavily doped Si and 29.1 nC/m for intrinsic Si. These values are significantly larger than the theoretically predicted ~1 nC/m [26,28,36]. Normalizing $\mu_{13}^{eff}$ by the bulk dielectric constant $\varepsilon_r$ [8,37,38] (see Section 4 in Supplemental Material [33]), we can further obtain the effective flexocoupling coefficient $f_{13}^{eff}$, which is found to be ~22000 V for heavily doped Si and ~320 V for intrinsic Si. These values are orders of magnitude larger than Kogan's prediction of 1~10 V for flexoelectric solids [8,39], and more specifically, ~3 V [28] and ~6 V [29] for Si.

We attribute the origin of the measured large flexoelectricity to the extrinsic barrier-layer mechanism [22,23], as Narvaez tentatively suggested in her thesis [30]. When silicon crystal contacts the Au electrode, electrons flow from silicon toward Au, leaving positively charged holes behind in the silicon and forming a Schottky barrier layer with a built-in electric field and a consequent interfacial polarization towards electrode (Fig. 1d) [40,41]. The formation of such barrier layer is verified by the anomalously large apparent dielectric constant $\varepsilon_r^{app}$ measured in Au/Si/Au capacitors (see Fig. S4a, in Supplemental Material [33]). Under bending deformation, the surface strain with opposite signs at two sides of the sample alters the initial interfacial polarization asymmetrically and results in a net flexoelectric polarization [22,23,32,42]. The basic idea is schematically illustrated in Fig. 1e and later theoretically formulated by equations, both of which depict the same barrier layer mechanism. We emphasize that this mechanism is different from the conventionally discussed surface contribution (or the mean-inner potential [43]) to flexoelectricity [29,44]. The former relies on the mesoscale barrier layer at the metal-semiconductor interface and yields anomalously large flexoelectric coefficients [22,23,32], whereas the latter originates from the spontaneous symmetry breaking of the unit-cell surface layer and exhibits a magnitude comparable with that of bulk flexoelectricity [29,44]. These two mechanisms may not be entirely independent, as the former has been considered as an amplified manifestation of the latter (Fig. 1e) due to the screening of polar discontinuity at the interface [22], but further conceptual distinction is peripheral to the scope of this paper. Within the barrier layer mechanism, higher carrier concentration causes larger interfacial polarization and consequently larger flexoelectric coefficient [22,23,32], which explains the observation that doped silicon is more flexoelectric than intrinsic silicon (Fig. 1c). It is worth mentioning that the observed dependence on doping concentration is stronger than that (40~80 nC/m) reported by Narvaez, which we suspect may be due to their use of a different type of dopant (phosphorus), a narrower doping range ($10^{12}$~$10^{17}$ cm$^{-3}$), and different electrode preparation procedures (from the vendor) [30].

*Contact author: xin.wen@icn2.cat

†Contact author: sshen@xjtu.edu.cn

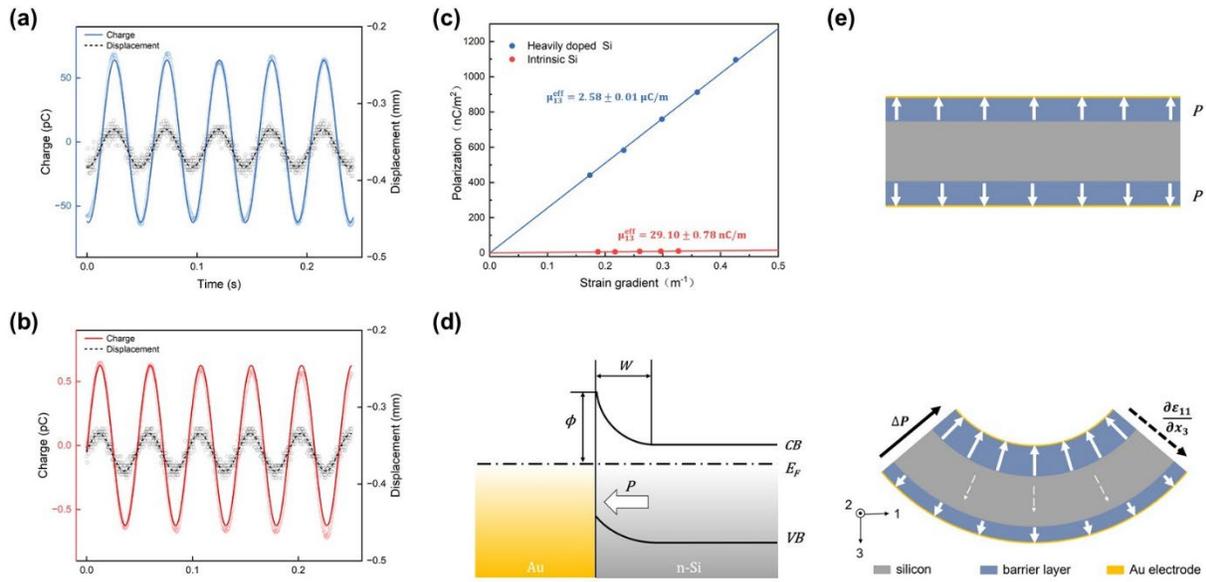

FIG 1. Effective flexoelectricity in silicon at room temperature. The Fourier-filtered first-harmonic displacement (dashed line) and charge (solid line) for heavily doped Si (a) and intrinsic Si (b). The open circles denote the raw data. (c) Flexoelectric polarization versus strain gradients of heavily doped Si and intrinsic Si. (d) Band-structure of the Schottky barrier between Si and Au electrode. $\Phi$ and $W$ represent the height and width of the Schottky barrier, respectively. CB, VB and $E_F$ represent the conduction band, valence band and Femi level, respectively. The white arrow represents the direction of the interfacial polarization $P$. (e) Schematic illustration of how the Schottky barrier layer contributes to flexoelectricity. The solid and dashed white arrows represent the interfacial polarization and the bulk flexoelectric polarization, respectively.

While the flexoelectric coefficient of heavily doped Si is nearly one hundred times larger than that of intrinsic Si at room temperature, we found that this disparity can be significantly modified by the variation of temperature. As temperature increases, as shown in Fig. 2a, the flexoelectric coefficient of intrinsic Si increases by approximately two orders of magnitude from -50 ℃ to 200 ℃, eventually merging into the curve for heavily doped Si that remains nearly constant over the same temperature range. Such distinct temperature dependences are mirrored by the capacitance measurement (Fig. 2b), showing a remarkable degradation of interfacial capacitance for intrinsic silicon as temperature decreases. Since both flexoelectricity and capacitance of Au/Si/Au are predominantly controlled by the Schottky barrier layer (Fig. 1, Fig. S4, S5), the divergent temperature-dependent behaviors between intrinsic and heavily doped Si indicates different degrees of thermal modifications to the barrier layer at the Si-Au interface. The main mediator in this process is presumably to be carrier concentration, the primary means of tuning semiconductor flexoelectricity [22,23,32,45,46].

To examine this hypothesis, we measured the evolution of carrier concentration ($N$) with temperature using the capacitance ($C$)-voltage ($V$) method for both intrinsic and doped Si samples (see Fig. S6, S7 in Supplemental Material [33]). Fig. 2c shows that the temperature dependence of carrier concentration resembles that of flexoelectricity and capacitance, showing large variation by seven orders of magnitude for intrinsic silicon but slight fluctuation within one order of magnitude for heavily doped Si from -50 ℃ to 200 ℃. Therefore, the charge carriers in intrinsic Si, once deficient at room temperature, is augmented as temperature increases, which, in turn, elevates its flexoelectric coefficient via the barrier layer mechanism (Fig. 2a). By the same token, the temperature-insensitive carrier concentration for doped Si is responsible for its temperature-insensitive flexoelectricity (Fig. 2a). Meanwhile, we notice that although the carrier concentration of intrinsic silicon monotonically increases up to 200 ℃, the flexoelectric coefficient get saturated at ~150 ℃ and does not further increase hand in hand with carrier concentration. Thus, there appear to be a polarity threshold at the Si-Au interface, potentially imposed by the tunneling current when the barrier layer gets narrower (see Fig. S8 in Supplemental Material [33]) or the dielectric degradation under high built-in field (see Fig. S9 of Supplemental Material [33]) [47–50]. Such flexoelectric saturation has been also reported in Nb-SrTiO$_3$ [32] and halide perovskites (MAPbBr$_3$, MAPbI$_3$) [23], and its exact origin warrants further investigation.

*Contact author: xin.wen@icn2.cat

†Contact author: sshen@xjtu.edu.cn

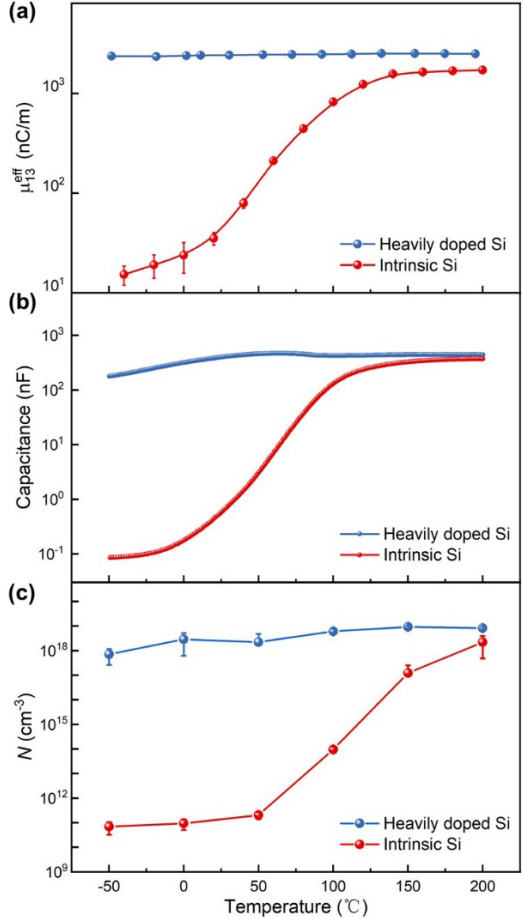

FIG 2. Temperature dependence of (a) the effective flexoelectric coefficient, (b) the capacitance of Au/Si/Au structure at 100kHz, and (c) the carrier concentration for heavily doped Si and intrinsic Si.

What renders intrinsic and doped silicon so distinct temperature dependences of free carriers? They have different sources of free carriers: intrinsic excitation and donor ionization (see Fig. S10 in Supplemental Material [33]). In intrinsic Si, the majority of free carriers arises from the thermal excitation of electrons from the valence band to the conduction band with a gap of 1.12 eV, which is of course determined by temperature. For heavily doped Si, in contrast, the majority of free carriers comes from the donor ionization that requires small ionization energy (~0.05 eV for As in Si [51]), which are predominantly determined by the doping concentration except at ultralow or ultrahigh temperatures.

To quantitatively capture the thermally excited flexoelectricity in intrinsic silicon, we reexamine the theory of semiconductor flexoelectricity. First, the effective flexoelectric coefficient of a bent semiconductor with a Schottky barrier layer can be expressed by the following equation [23,32]

$$\mu_{13}^{eff} = \sqrt{\frac{N\varepsilon_0\varepsilon_r}{2\phi}}\frac{qt\varphi}{2}, \quad (1)$$

where $N$ is the carrier concentration of the semiconductor, $\varepsilon_0$ is the vacuum permittivity, $\varepsilon_r$ is the relative dielectric constant, $\phi$ is the height of the Schottky barrier, $q$ is the electronic charge, $t$ is the thickness of the sample, and $\varphi$ is the surface deformation potential that denotes the change in barrier height $\phi$ due to surface strain.

Now, we introduce temperature dependence into Eq. (1). The carrier concentration due to intrinsic excitation can be expressed as [40]

$$N = \sqrt{N_C N_V}\, e^{-\frac{E_g}{2k_B T}}, \quad (2)$$

where $N_C$ and $N_V$ represent the effective density of states in the conduction and valence bands, respectively, $E_g$ is the bandgap of semiconductor, $k_B$ is the Boltzmann constant (1.38×10$^{-23}$ J/K), and $T$ represent the absolute temperature. Substituting Eq. (2) into Eq. (1), we obtain the temperature-dependent effective flexoelectric coefficient:

$$\mu_{13}^{eff} = \varphi t (N_C N_V)^{\frac{1}{4}}\left(\frac{\varepsilon_0\varepsilon_r q}{2\phi}\right)^{\frac{1}{2}} e^{-\frac{E_g}{4k_B T}}. \quad (3)$$

Differentiate with respect to the reciprocal of temperature, and neglecting the weak temperature independence of some parameters (see Section 9 in Supplemental Material [33]), the seemingly complex Eq. (3) can be distilled as an explicit and concise relationship between the bandgap and the flexoelectric coefficient

$$\frac{d(\ln(\mu_{13}^{eff}))}{d(T^{-1})} \approx -\frac{E_g}{4k_B T} \quad (4)$$

*Contact author: xin.wen@icn2.cat

†Contact author: sshen@xjtu.edu.cn

To verify this connection, we plot the relationship of $\ln(\mu_{13}^{eff})$ and $T^{-1}$ based on multiple measurements of intrinsic silicon (Fig. 3a). The slopes of $\ln(\mu_{13}^{eff})$ versus $T^{-1}$ consistently yield a $E_g$ of ~1.12 eV (Fig. 3b), which is precisely the bandgap of Si at room temperature [40]. Therefore, despite the qualitative nature of the expression for $\mu_{13}^{eff}$ in Eq. (1) [23], we show here that its temperature dependence can be quantitatively correlated with the band gap $E_g$. While Fig.3a also shows appreciable intersample variations in magnitude for intrinsic Si (absent in doped Si, see Fig. S12 in Supplemental Material [33]), presumably due to the stochasticity of the surface states [52], we found that an annealing process before the measurements can improve this issue while the linearly fitted $E_g$ remains unchanged as ~1.12 eV (see Fig. S13, S14 in Supplemental Material [33]). This consistency indisputably identifies the intrinsic excitation of free carriers as the origin of the thermally enhanced flexoelectricity in intrinsic silicon. Similarly, the bandgap of intrinsic germanium (Ge), another prototype material of first-generation semiconductors, can also be reproduced in the same manner (Fig. 3b, Fig. S15 in Supplemental Material [33]), demonstrating the universality of Eq. (4).

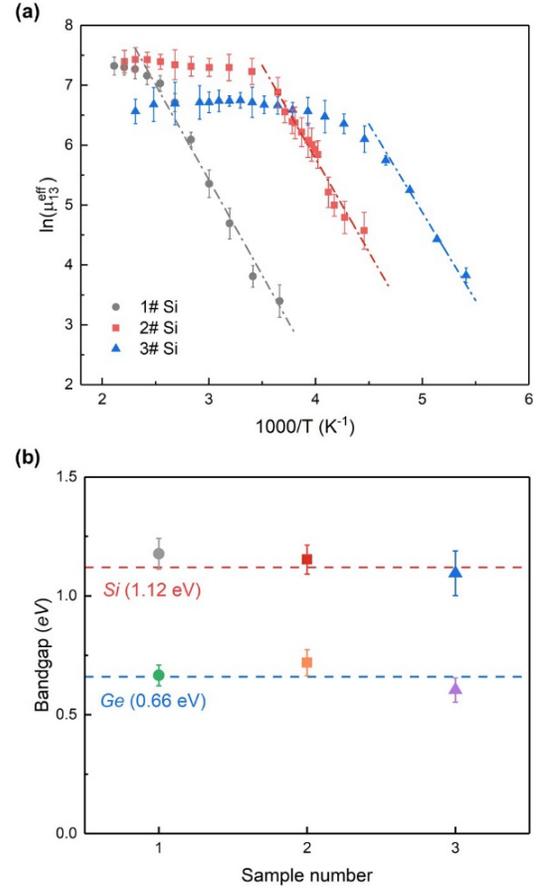

FIG 3. (a) The fitting of the flexoelectric coefficient for different intrinsic Si samples. (b) The comparison between the fitted bandgap (symbols) and the textbook bandgap (dashed line) of Si and Ge.

In conclusion, we show that the flexoelectric coefficient of Si can vary by up to two orders of magnitude depending on the carrier concentration that can be modified either by chemical doping or by intrinsic excitation of free carriers. The latter, in particular, highlights the critical role of temperature—an often overlooked factor in prior studies [11–21,53–58]—in modulating the flexoelectricity in semiconductors, depicted by our proposed theoretical relationship among bandgap, temperature and flexoelectricity. Furthermore, the reported flexoelectric coefficient of silicon is comparable with that of some ferroelectric ceramics [7], suggesting tantalizing potential for integrating flexoelectric silicon devices into its readily compatible silicon-based technology.

*Contact author: xin.wen@icn2.cat

†Contact author: sshen@xjtu.edu.cn

## ACKNOWLEDGMENTS

We thank J. Liu for her code to process the experimental data. This work was supported by the 111 Project (No. B18040) and the National Natural Science Foundation of China (No. 123B2027).

*Contact author: xin.wen@icn2.cat

†Contact author: sshen@xjtu.edu.cn

*Contact author: xin.wen@icn2.cat

†Contact author: sshen@xjtu.edu.cn

*Contact author: xin.wen@icn2.cat

†Contact author: sshen@xjtu.edu.cn